\documentclass{svjour2}                    % onecolumn
\smartqed  % flush right qed marks, e.g. at end of proof
\usepackage{amsmath,bm}
\usepackage{graphicx}
\usepackage{acronym}
\usepackage{color}
\usepackage[normalem]{ulem} % to cross out text with \sout{}

\graphicspath{{./GRAFICAS/}{./}}

\definecolor{darkBlue}{rgb}{0,0,0.6}
\definecolor{darkRed}{rgb}{0.5,0,0}
\definecolor{darkGreen}{rgb}{0,0.5,0}

\newcommand{\Tr}{\mathrm{Tr}}
\newcommand{\I}{\mathrm{I}}
\newcommand{\s}{\mathrm{s}}
\newcommand{\sj}{\mathrm{sj}}
\newcommand{\runs}{\mathrm{runs}}

\newcommand{\MAX}{\mathrm{MAX}}

\newcommand{\rz}{\hat r^{(0)}}
\newcommand{\rk}{\hat r^{(k)}}
\newcommand{\rkp}{\hat r^{(k')}}

\newcommand{\rka}{{\rk_\alpha}}
\newcommand{\rkb}{{\rk_\beta}}
\newcommand{\rkc}{{\rk_\gamma}}
\newcommand{\rkd}{{\rk_\delta}}
\newcommand{\rkx}{{\rk_x}}
\newcommand{\rky}{{\rk_y}}
\newcommand{\rkz}{{\rk_z}}
\newcommand{\rza}{{\rz_\alpha}}
\newcommand{\rzb}{{\rz_\beta}}

\newcommand{\rzx}{{\rz_x}}
\newcommand{\rzy}{{\rz_y}}
\newcommand{\rzz}{{\rz_z}}
\newcommand{\bk}{b^{(k)}}
\newcommand{\bkp}{b^{(k')}}
\newcommand{\bz}{b^{(0)}}
\newcommand{\xk}{\vec x^{(k)}}

\newcommand{\xz}{\vec x^{(0)}}

\newcommand{\Nb}{N_b}
\newcommand{\ab}{{\alpha\beta}}
\newcommand{\cd}{{\gamma\delta}}
\newcommand{\abcd}{{\alpha\beta\gamma\delta}}

\newcommand{\Cxy}{C^{(\sigma)}_{xy}(\vec x)}
\newcommand{\Cxz}{C^{(\sigma)}_{xz}(\vec x)}
\newcommand{\Cabcd}{C^{(\sigma)}_{\alpha\beta\gamma\delta}(\vec x)}

\newcommand{\C}{C^{(\sigma)}(\vec x)}
\newcommand{\Cp}{C^{(p)}(\vec x)}

\newcommand{\Celas}{\mathcal{C}}

\begin{document}

\title{Emergent $SO(3)$ Symmetry of the Frictionless Shear Jamming Transition
% \thanks{Grants}
}
% \subtitle{Do you have a subtitle?\\ If so, write it here}

\titlerunning{Frictionless Shear Jamming}        % if too long for running head

\author{
Marco Baity-Jesi \and
Carl P. Goodrich \and
Andrea J. Liu \and
Sidney R. Nagel\and
James P. Sethna
}


\institute{
Marco Baity-Jesi \at Institut de Physique Th\'eorique, DRF, CEA Saclay, France; and Department of Physics and Astronomy, University of Pennsylvania, Philadelphia, Pennsylvania 19104, USA
              \email{marco.baity-jesi@cea.fr}           %  \\
           \and
Carl P. Goodrich          \at School of Engineering and Applied Sciences, Harvard University, Cambridge, MA 02138, USA; and Department of Physics and Astronomy, University of Pennsylvania, Philadelphia, Pennsylvania 19104, USA
                     \and
Andrea J. Liu          \at Department of Physics and Astronomy, University of Pennsylvania, Philadelphia, Pennsylvania 19104, USA
                     \and
Sidney R. Nagel          \at James Franck Institute, Enrico Fermi Institute, and Department of Physics, The University of Chicago, USA
                     \and
James P. Sethna          \at Department of Physics, Cornell University, Ithaca, New York 14850, USA
          }

\maketitle

\begin{abstract}
We study the shear jamming of athermal frictionless soft spheres, and find that in the thermodynamic limit, a shear-jammed state exists with different elastic properties from the isotropically-jammed state. 
For example, shear-jammed states can have a non-zero residual shear stress in the thermodynamic limit that arises from long-range stress-stress correlations. As a result, the ratio of the shear and 
bulk moduli, which in isotropically-jammed systems vanishes as the jamming transition is approached from above, instead approaches a constant. Despite these striking differences, we argue that in a 
deeper sense, the shear jamming and isotropic jamming transitions actually have the same symmetry, and that the differences can be fully understood by rotating the six-dimensional basis of the elastic modulus tensor. 
\end{abstract}

%%%%%%%%%%%%%%
%Introduction%
%%%%%%%%%%%%%%
\section{Introduction}

The critical jamming transition of soft frictionless spheres at zero temperature provides a framework for understanding the mechanical and low-temperature response of a wide range of disordered 
materials~\cite{liu:10,xu:07,wyart:08,phillips79,phillips:81,boolchand:05,song:08,henkes:10,papanikolaou:13,zhang:09,donev:04,zeravcic:09,mailman:09,goodrich:14c,somfai:07,shundyak:07}. However, 
in order to connect the jamming scenario to systems that include, for example, temperature, attractive interactions, or friction, it is important to understand how these effects perturb the physics 
of the jammed solid. For example, spheres with frictional interactions are able to jam at lower densities than those without friction when subjected to an external shear stress, leading to a so-called 
anisotropic ``shear-jammed" state~\cite{bi:11}.

Several studies have examined shear jamming in frictionless particle packings~\cite{bertrand:16,imole:13,vinutha:16,vinutha:16b,roux:08a,roux:08b} and have shown that finite systems can shear 
jam~\cite{bertrand:16,imole:13}.
Here, we focus primarily on mechanical properties of the shear-jammed state of {three-dimensional} athermal, frictionless, soft repulsive spheres, and report three findings. 

First,  although the density range over which frictionless shear jamming occurs vanishes in the thermodynamic limit, as previously observed~\cite{bertrand:16}, frictionless shear jamming is \emph{not} 
a finite-size effect.  In isotropic jamming, the lack of long-range correlations in the bond forces associated with particle-particle contacts means that the average residual shear stress, $s$, is 
zero by symmetry and fluctuations scale as $N^{-1/2}$ relative to the pressure, $p$. However, because the shear-jammed state has long-ranged bond force correlations, it can support a non-zero 
residual shear stress as well as pressure in the thermodynamic limit, unlike the isotropically-jammed state, which supports only pressure. 
{States of this kind had already shown to exist in the jammed phase \cite{roux:08a}. Through a careful scaling analysis we show that such states can support a shear stress even infinitesimally above the shear-jamming transition in the thermodynamic limit.}

Second, the shear-jamming transition is different from the isotropic jamming transition. In isotropic jamming, the property that shear stress fluctuations vanish in the thermodynamic limit gives 
rise to different scalings of the bulk modulus, $B$, and shear modulus, $G$~\cite{goodrich:16}; the vanishing of $G/B$ as the transition is approached from above is one of the 
defining characteristics of jamming. In shear jamming, the fact that the shear stress does not vanish in the thermodynamic limit changes the scaling of the shear modulus so that $G/B$ remains 
constant at the shear-jammed transition, as has been argued in various contexts \cite{wyart:05,wyart:11,zaccone:14}.

Our third and main finding is that, despite these striking differences, the shear and isotropic jamming transitions share an important symmetry.  The effect of shear on the scaling behaviors discussed 
above is a result of an induced rotation of the six-dimensional eigenvectors of the stiffness tensor~\cite{roux:08b}. Once this rotation is taken into account, the $SO(3)$ symmetry of isotropic 
jamming is preserved in the rotated subspace. Therefore, despite having different scaling exponents for the shear modulus, shear jamming and isotropic jamming should be considered members of the 
same universality class.

%%%%%%%%%%%%%%%%%%%%%%%%%%%%%%%%%%%%%%%%%%
%Model, simulations and basic observables%
%%%%%%%%%%%%%%%%%%%%%%%%%%%%%%%%%%%%%%%%%%
\section{Systems Studied}
We study 3$d$ athermal disordered packings of $N=64$ to $N=4096$ frictionless spheres in cubic boxes with periodic boundary conditions. Particles $i$ and $j$, of radius $R_i$ and $R_j$ and whose centers are separated by a distance $r_{ij}$, interact through the potential
\begin{equation}\label{eq:U}
 U(r_{ij}) = \frac{U_0}{\alpha} \left( 1 - \frac{r_{ij}}{R_i+R_j}\right)^\alpha \Theta\left( 1 - \frac{r_{ij}}{R_i+R_j}\right)\,.
\end{equation}
We show results for $\alpha=2$, corresponding to harmonic repulsions between particles. 
We also carried out simulations for the Hertzian potential ($\alpha=5/2$), and found no difference beyond the expected ones found in isotropic jamming%
\footnote{A rescaling is necessary, as shown, e.g., in reference \cite{vitelli:10}.}.
Here $U_0$ sets the energy scale, and $\Theta(x)$ is the Heaviside step function. 
The packing fraction is $\phi=\sum_i V_i/V$, where $V_i$ is the volume of particle $i$ and $V$ is the volume. In order to avoid crystallization, we study 50:50 bidisperse 
packings with 1:1.4 diameter ratio.

\section{Creating Shear Jammed Packings}
To create our packings, we conduct $N_\runs = $ 1000 to 10000 independent runs for each value of $\phi$ and $N$ studied. 
Each run begins with a completely random (infinite-temperature) configuration. We minimize the total potential energy [the sum of the pair interactions of Eq.~\eqref{eq:U}] to a nearby local minimum with the FIRE algorithm \cite{bitzek:06}.
% \footnote{With a precision $\hat\varepsilon=10^{-12}$.}
At this point, a fraction $f_\I(\phi)$ of the systems are isotropically-jammed, as in Ref.~\cite{ohern:03}.%
\footnote{We establish that a configuration is jammed when it has a rigid backbone (not all the particles are rattlers) and the number of contacts in the rigid cluster is above the isostatic value.} 
We focus on the remaining $N_\runs (1-f_\I(\phi) )$ unjammed configurations. Using strain steps of $\delta \gamma = 0.02$, we apply simple shear in the $xy$ direction at 
constant packing fraction, minimizing the energy after each step, until the system either jams or the strain exceeds a threshold of $\gamma_\MAX=0.4$. The fraction of states that 
were initially unjammed but jammed due to shear strain 
is denoted by $f_\s$; $f_\sj = (1-f_\I) f_\s$ is the fraction out of all states that are shear-jammed.

\begin{figure}[!tbh]
\centering
\includegraphics[width=.8\columnwidth]{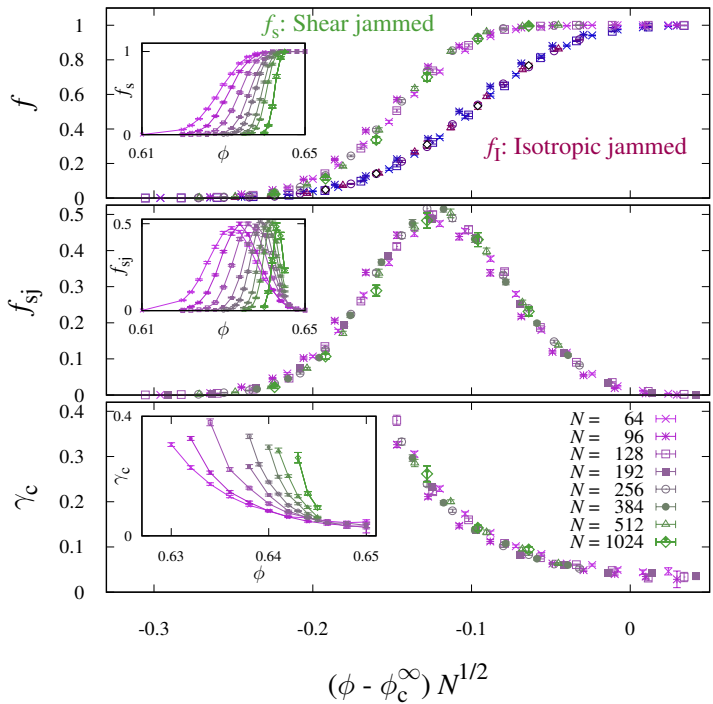}
\caption{(a) The fraction, $f_\I$, of states that isotropically jam and the fraction $f_\s$ of states that do not isotropically jam but shear jam, respectively, as functions of $(\phi-\phi_\mathrm{c}^\infty)\sqrt{N}$.  In all panels of this figure, $\phi_\mathrm{c}^\infty = 0.6470(5)$.  Note that
the same choice of the scaling variable collapses both curves, though the scaling function is different and may depend on the protocol. 
In the inset we show uncollapsed data, $f_\s(\phi)$ with system sizes as indicated in the legend to part (c) of the figure.
Throughout this paper, we use a red-blue scale (for increasing system size) for regular jammed packings, 
and violet-green scale for shear-jammed systems. (b) The fraction $f_\sj$ as a function of the scaling variable $(\phi-\phi_\mathrm{c}^\infty)\sqrt{N}$ of packings that shear jam relative to the total number of packings.  The inset depicts $f_\sj(\phi)$ for different values of $N$. (c) Median of the strain $\gamma_\mathrm{c}$ needed to shear jam the system. In the main plot we show a collapse as a function of the scaling variable
$(\phi-\phi_\mathrm{c})\sqrt{N}$. The strain $\gamma_\mathrm{c}$ does not need to be rescaled by $N$ in order to collapse the data, implying that the shear-jammed state exists in the thermodynamic limit and is distinct from the isotropically-jammed state. The uncollapsed data is shown in the inset.}
\label{fig:f}
\end{figure}

For isotropic jamming, the fraction of jammed states, $f_\I(\phi)$, collapses for different system sizes $N$ when 
plotted against $(\phi-\phi_\mathrm{c}^\infty)\sqrt{N}$~\cite{ohern:03}.  Thus, $f_\I(\phi)$ approaches a step function centered around $\phi_\mathrm{c}^\infty$ in the thermodynamic limit. 
We see in Fig.~\ref{fig:f}(a) that $f_\s$ also collapses with the same scaling variable and with the same $\phi_\mathrm{c}^\infty=0.6470(5)$.  
Fig.~\ref{fig:f}(a) shows that for any system size, the onset of shear jamming lies below the onset of isotropic jamming, consistent with earlier results~\cite{imole:13,madadi:04}, and that the difference between the two packing 
fractions appears to vanish in the thermodynamic limit, consistent with Ref.~\cite{bertrand:16}. 

Figure~\ref{fig:f}(b) shows the fraction of shear-jammed packings relative to the total number of packings, $f_\sj=(1-f_\I)f_\s$.  The curve is bell-shaped due to the presence of two effects. At low $\phi$ few packings shear jam because the packing
density is too low. At high $\phi$ few packings are shear jammed because they jam even without shear.
The peak of $f_\sj(\phi)$ indicates the optimal packing fraction to obtain shear-jammed packings through our protocol. Note that the curves for different system sizes $N$ collapse with $(\phi-\phi_\mathrm{c}^\infty)\sqrt{N}$, as for $f_\I$ and $f_\s$.

Figure~\ref{fig:f}(c) shows the median strain $\gamma_\mathrm{c}$ at which an initially unjammed configuration jams due to shear.%
\footnote{Since we apply strains only up to $\gamma_\MAX=0.4$, we are unable to capture the high $\gamma$ tail of the distribution of $\gamma_\mathrm{c}^\Lambda$. As a result, we show the median of 
$\gamma_\mathrm{c}^\Lambda$ instead of the mean, and consider only values of $\phi$ and $N$ such that $f_\s > 0.5$.}
From the inset to Figure~\ref{fig:f}(c), we see that systems at lower values of $\phi$ shear-jam at higher values of $\gamma_\mathrm{c}$, consistent with earlier results~\cite{bertrand:16}. Each curve 
ends at low $\phi$ or high $\gamma_\mathrm{c}$ due to statistics; the more 
configurations we study, the higher each curve extends in $\gamma_\mathrm{c}$.  Note that our results collapse for different system sizes in the same fashion as the fraction of shear-jammed 
states, {\it i.e.} as $(\phi-\phi_\mathrm{c}^\infty)\sqrt{N}$, without requiring any scaling of $\gamma_c$ with $N$. 
To understand the significance of this result, consider choosing $\phi$ for each $N$ such that $(\phi-\phi_c^\infty) N^{1/2}$ is fixed, and asking what happens with increasing $N$.  If the $y$-axis in Fig.~\ref{fig:f}(c) had been $\gamma_c N^x$ where $x>0$, then we would find that the strain needed to shear-jam the system would vanish as $N \rightarrow \infty$.  In that case, the properties of the shear-jammed state would be identical to those of the isotropically-jammed state in the thermodynamic limit.  If, on the other hand, we had needed to collapse $\gamma_c N^x$ with $x<0$, the strain required to shear-jam the system would have diverged as $N \rightarrow \infty$, implying that shear-jamming is not possible in the thermodynamic limit.  The fact that no power of $N$ was needed to collapse the data with $\gamma_c$ shows that shear-jamming is possible in the thermodynamic limit even though it occurs at $\phi_c^\infty$, and that the properties of the shear-jammed state are different from those of the isotropically-jammed state.

% and have the same properties even in the thermodynamic limit, though they become harder to reach with our protocol.

%
%Note that each SJ configuration $\Lambda$ jams at a certain strain, $\gamma_\mathrm{c}^\Lambda$.  Since we only apply strains up to $\gamma_\MAX=0.4$, the high-strain 
%tail of the distribution of $\gamma_\mathrm{c}^\Lambda$ is truncated.  To avoid bias due to this truncation, we consider the median value of $\gamma_\mathrm{c}$ instead of 
%the mean, and consider only values of $\phi$ and $N$ such that at least half of the measured values of $\gamma_\mathrm{c}^\Lambda$ are smaller reduce our  than $\gamma_\MAX$. 
%Note that in the presence of large-tail distributions the median is often a better descriptor than the average \cite{medians:14}.
 
%Fig.~\ref{fig:gammac} shows that systems at lower values of $\phi$ shear-jam at higher values of $\gamma_\mathrm{c}$, consistent with earlier results~\cite{bertrand:16}.  

%%%%%%%%%%%%%%%%%%%%%%%
%NEW SCALING EXPONENTS%
%%%%%%%%%%%%%%%%%%%%%%%
\section{Mechanical Properties of the Shear Jammed State}

We now consider systems strained past the point of shear-jamming into the jammed state. Specifically, we generate systems at logarithmically spaced values of shear stress $\sigma_{xy}$. 
To obtain these configurations, we start with a system that has been sheared beyond the onset of shear jamming as described above. We then pick a target shear stress, $\sigma_{xy}^\mathrm{t}$ at 
the upper end of the range of shear stresses we want to study,
and adjust the strain until the target shear stress is obtained to within $1\%$. 
To approach  $\sigma_{xy}^\mathrm{t}$ efficiently we exploit the definition of the elastic constants, $C_{xyxy}=\frac{d\sigma_{xy}}{d\gamma_{xy}}$.  To linear order the needed strain is
$\frac{\sigma_{xy}^\mathrm{t} - \sigma_{xy}}{C_{xyxy}}$. 
We combine this with the Newton-Raphson method to tune the system to $\sigma_{xy}^\mathrm{t}$. By iteratively lowering $\sigma_{xy}^\mathrm{t}$, we create configurations with $\sigma_{xy}$ spanning 
many orders of magnitude. This protocol is very similar to that used in isotropic jamming to obtain systems at target pressures~\cite{goodrich:16,goodrich:12,goodrich:14}. 
{Although some dependence on the packing fraction is expected, we find that it is weak for the system sizes studied.  As a result, all the runs shown are for $\phi=0.643$, except for $N$=4096, for which $\phi=0.645$. 
\footnote{{For $\phi=0.643$, $N=4096$, the data yield consistent results, but with larger error bars due to the difficulty in obtaining shear-jammed states 
(see figure \ref{fig:f}--center,inset).}}}

\begin{figure}[!tb]
 \includegraphics[width=\columnwidth]{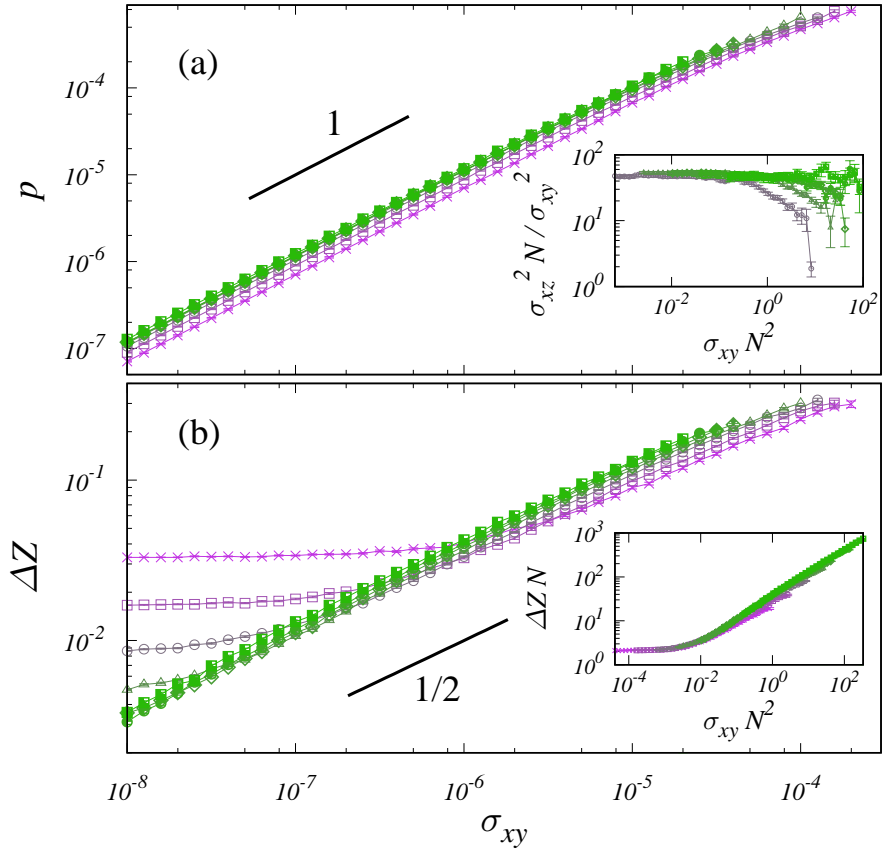}
 \caption{
 (a) The pressure $p$ is proportional to the shear stress $\sigma_{xy}$ for all system sizes. In the inset the collapse shows that stress orthogonal to the direction of shear jamming, $\sigma_{xz}$, obeys the 
 scaling relation $\sigma_{xz}^2\sim\sigma_{xy}^2/N$. 
 (b) The excess contact number $\Delta Z$ as a function of shear in the direction of shear jamming, $\sigma_{xy}$, for several system sizes $N$. The inset shows a collapse for the same data set. The scaling variable for the stress is $\sigma_{xy}N^2$.
  }
 \label{fig:dZ-p}
\end{figure}

Figure~\ref{fig:dZ-p}(a) shows that the pressure $p$ scales linearly with $\sigma_{xy}$ with a prefactor that is independent of $N$ in the large $N$ limit, 
{ reminiscent of the scaling observed in dense granular flows \cite{roux:08a}.  In that case, the shear stress and pressure approach nonzero values as the shear rate is reduced to zero; here, we are exploring the behavior near the shear-jamming transition, where the shear stress and pressure both vanish. } We note that the scaling $\sigma_{xy} \sim p$ contrasts with the scaling observed in 
isotropic jamming where the shear stress fluctuates around zero, with the magnitude of fluctuations scaling as $N^{-1/2}$~\cite{goodrich:14}, such that $\sigma_{xy}^2 \sim p^2/N$~\cite{goodrich:16}.  
%This scaling follows from the behavior of bond correlation functions~\cite{goodrich:16}: $\sigma_{xy}^2$ and $p^2$ can be written in terms of integrals over bond correlation functions that are short-ranged for $\sigma_{xy}^2$ and long-ranged for $p^2$.  In the shear-jammed case, the bond correlations are long-ranged not only for $p^2$ but also for $\sigma_{xy}^2$ (see Appendix), reflecting the shear-induced anisotropy in the packing structure. This leads to the scaling $p \sim \sigma_{xy}$ shown in Fig.~\ref{fig:dZ-p}(a). 
The inset to Figure \ref{fig:dZ-p}(a) shows that the shear stress in orthogonal directions ({\it e.g.} $\sigma_{xz}$) vanishes as $1/\sqrt{N}$ in the thermodynamic limit.

Figure~\ref{fig:dZ-p}(b) shows the behavior of the excess contact number ($\Delta Z \equiv Z - Z_\mathrm{iso}$, where $Z_\mathrm{iso} \equiv 2d - 2d/N$). 
Just as for isotropic jamming, $\Delta Z$ approaches $2/N$ as expected from counting arguments~\cite{goodrich:12} in the limit $\sigma_{xy} \rightarrow 0^+$. The inset shows that scaling 
collapse takes the form $\Delta Z N = F(\sigma_{xy}N^2)$, in analogy to the form $\Delta Z N = F_I(p N^2)$ observed in isotropic jamming.  Thus, at large $\sigma_{xy} N^2$, 
$\Delta Z\sim \sigma_{xy}^{1/2} \sim p^{1/2}$.  Note that for isotropic jamming, $\Delta Z \sim p^{1/2}$ at high $p N^2$~\cite{goodrich:12} but $\Delta Z \sim N^{1/4} \sigma_{xy}^{1/2}$ in
that limit, since $\sigma_{xy} \sim N^{-1/2}$.

%system is isostatic with a finite-size correction as found in Ref.~\cite{goodrich:12}. %Once that correction is subtracted off, 
%Fig.~\ref{fig:dZ-p}(a) shows that for our SJ packings, the excess contact 
%number scales as $\Delta Z\sim \sigma_{xy}^{1/2}$ at large $\sigma_{xy}$, again consistent with the isotropic result that $\Delta Z \sim p^{1/2}$ at large $p$.

%Note that in isotropically-jammed packings, the average stress is zero by symmetry and fluctuations vanish as $1/\sqrt{N}$~\cite{goodrich:16}. In SJ packings, we 
%find that the scaling behavior depends on the direction of the applied strain.  The average stress in the direction of the applied strain, $\sigma_{xy}$, scales linearly 
%with pressure (Fig.~\ref{fig:dZ-p}(b)), which is consistent with the presence of macroscopic friction in sheared systems \cite{roux:08a}.
%By contrast, the average value of $\sigma_{xz}$ vanishes by symmetry, as in the isotropically-jammed case (Fig.~\ref{fig:dZ-p}(b, inset)).  

Similarly, the scaling behavior of the elastic constants reflects the anisotropy of the shear-jammed state. In the thermodynamic limit, isotropically-jammed systems are characterized by just 
two elastic constants, the bulk modulus $B$ (which jumps discontinuously at jamming, scaling as $B\sim\Delta Z^{\gamma_\mathrm{B}}$ where $\gamma_\mathrm{B}=0$) and the shear modulus $G$ 
(which increases continuously, scaling as  $G\sim\Delta Z^{\gamma_\mathrm{G}}$ with $\gamma_\mathrm{G}=1$).  
For shear-jammed systems, the shear modulus depends on the direction in which it is measured; in addition to the bulk modulus, we characterize the elasticity by the response to shear that is 
parallel ($C_{xyxy}$) and perpendicular ($C_{xzxz}$) to the initially imposed strain. 

\begin{figure}[!tbh]
\centering
\includegraphics[width=0.8\columnwidth]{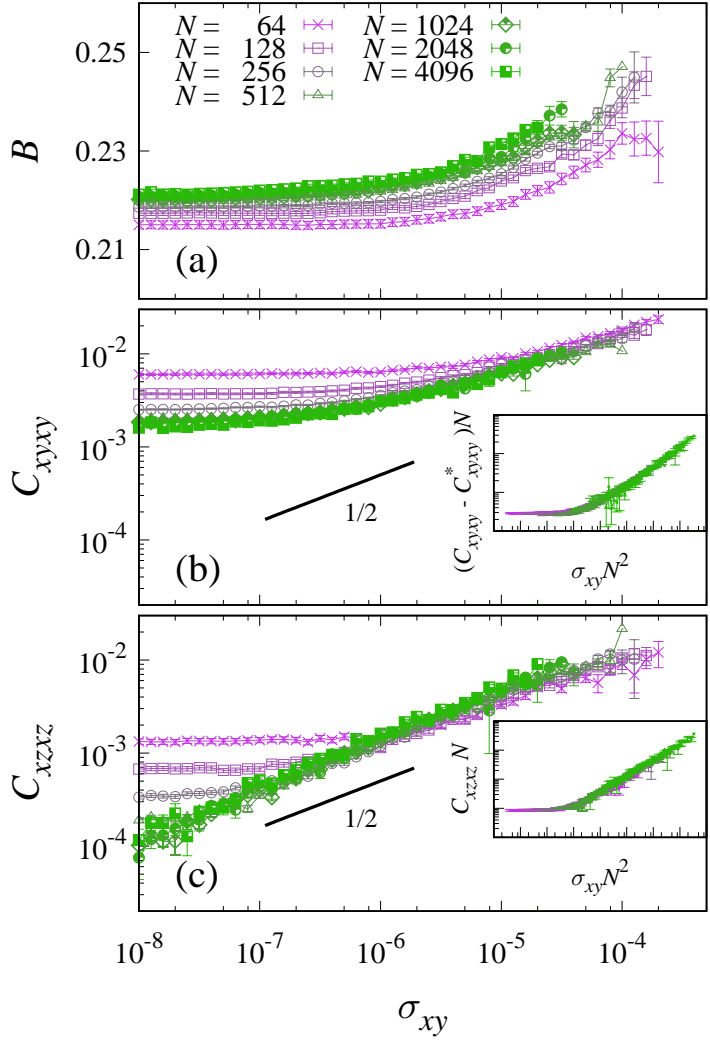}
\caption{Scalings of elastic moduli for the shear-jammed state as a function of the stress in the shearing direction, $\sigma_{xy}$. 
(a) The bulk modulus $B$ approaches a nonzero value, $B \rightarrow B_0 =0.221(1)$, as $\sigma_{xy}$ approaches the shear-jamming transition from above. 
(b) The elastic constant $C_{xyxy}$ approaches a nonzero value, $C_{xyxy}^*=0.0015(1)$ as 
$\sigma_{xy} \rightarrow 0^+$. 
% The dark black line shows the power law $\sigma^{1/2}$, corresponding to the corrections to scaling at large $\sigma_{xy}$ that we predict through the correlations (see Appendix). 
The inset shows the 
scaling collapse of $C_{xyxy}-C_{xyxy}^*$.  (c) The elastic constant $C_{xzxz}$ vanishes as $C_{xzxz} \sim \sigma_{xy}$ as $\sigma_{xy} \rightarrow 0^+$. 
The inset shows the scaling collapse of $C_{xzxz}$.}
\label{fig:G}
\end{figure}

Figure~\ref{fig:G}(a) shows that the bulk modulus approaches a constant as $\sigma_{xy} \rightarrow 0$ (or equivalently in the thermodynamic limit, as $\Delta Z \rightarrow 0$), just as for 
isotropic jamming. However, Fig.~\ref{fig:G}(b) shows that $C_{xyxy}$ also approaches a constant in that limit (albeit a much smaller constant than for the bulk modulus). This scaling of 
$C_{xyxy}$ represents a striking departure from the shear modulus of isotropic jamming, where $C_{xyxy}$ vanishes at the transition. This difference persists even in the thermodynamic limit, 
showing that the anisotropic shear-jammed state remains distinct from the isotropically-jammed state in that limit. 
\footnote{{Despite the stressed differences between isotropic and shear jamming, we will show in section \ref{sec:emergent} that it is possible to rationalize the different exponents 
in order to show that the shear does \emph{not} change the universality class of the jamming transition.}}
Note, however, that Figure~\ref{fig:G}(c) shows that the orthogonal shear 
modulus $C_{xzxz}$ scales identically to the shear modulus in the isotropic case: $C_{xzxz} \sim  \sigma_{xy}^{1/2} \sim \Delta Z$ with a finite-size plateau at low $\sigma_{xy}$ that scales 
as $N^{-1}$ (inset)~\cite{goodrich:12}.

We can understand the results of Figures~\ref{fig:dZ-p} and \ref{fig:G} by first reviewing the arguments of Ref.~\cite{goodrich:16} for isotropic jamming. Because long-range order in the orientation 
and magnitude of the contact forces is absent in isotropically-jammed packings, the average deviatoric stress vanishes and its fluctuations scale as $s^2 \sim p^2/N$. The $N$-dependence in this relation 
leads to different scalings, $s \sim \Delta Z^{\delta_\sigma}$ and $p \sim \Delta Z^{\delta_p}$ for the shear stress and pressure with respect to $\Delta Z$:  Ref.~\cite{goodrich:16} shows that 
$\delta_\sigma - \delta_p = \psi/2$, where $\psi=1$ is the finite-size scaling exponent. Because the moduli are derivatives of the stresses with respect to the strains, this scaling relation leads to 
$\gamma_G = \gamma_B + 2(\delta_\sigma - \delta_p)$ where $\gamma_G$ and $\gamma_B$ are the scaling exponents for the shear and bulk moduli. Thus, the lack of long-range orientational order leads directly 
to the difference in $\delta_\sigma$ and $\delta_p$, which in turn leads directly to the difference in $\gamma_G$ and $\gamma_B$. 

In shear-jammed systems, by contrast, Figure~\ref{fig:dZ-p} shows that $p \sim \sigma_{xy}$ with an $N$-independent prefactor.  As a result, $\delta_{\sigma_{xy}}=\delta_p$ and the exponent $\gamma_{xyxy}$ 
for $C_{xyxy}$ is identical to $\gamma_B$ for the bulk modulus, as we indeed find in Figure~\ref{fig:G}.  On the other hand, the inset to Figure~\ref{fig:dZ-p}(a) shows that $\sigma_{xz}^2 \sim \sigma_{xy}^2/N$, 
implying $\delta_{\sigma_{xz}}=\delta_p+\psi/2$, as in the isotropic jamming case, leading to different exponents for $C_{xzxz}$ and $B$, as we find in Figure~\ref{fig:G}.

As we show in the Appendix, these scalings can be derived from the behavior of bond correlation functions introduced in Ref.~\cite{goodrich:16}. Integrals over these correlation functions contribute to 
components of the stress tensor.  If a bond correlation function approaches a constant value at large separations, then the corresponding component of the stress tensor remain constant as $N \rightarrow \infty$; 
if it is short-ranged, the corresponding component of the stress tensor vanishes as $1/\sqrt{N}$.  In the Appendix, we show that these bond correlation functions can be written in a way that respects the broken 
symmetry of shear-jammed systems. In such systems, there is still no long-ranged order in the $xz$ orientation, but { shear-jamming gives rise to long-ranged bond-force correlations so that long-ranged order arises in the $xy$ orientation. As a result, the system can support a nonzero shear stress in the thermodynamic limit, implying $\delta_{\sigma_{xy}}=\delta_p$ and hence} the results shown 
in Figure~\ref{fig:G}.

In the limit where a vanishing strain $\gamma_c$ is applied to shear-jam the system, one might expect the properties of the shear-jammed system to become identical to those of an isotropically-jammed system.  
In particular, $C_{xyxy}$ approaches a constant, $C_{xyxy}^*$ for shear-jamming, but vanishes for isotropic jamming.  We observe that $C_{xyxy}^*$ decreases with $\gamma_c$, consistent with it vanishing 
as $\gamma_\mathrm{c}$ goes to zero (not shown).
\section{Emergent Symmetry at the Shear Jamming Transition}\label{sec:emergent}

A remarkable symmetry emerges at the shear
jamming transition.  We first note an important symmetry at the isotropic jamming transition.
In the thermodynamic limit, the bulk modulus
jumps to a non-zero value while the shear modulus remains zero, so
$C_\abcd = B \delta_\ab \delta_\cd$. By spatial isotropy, the various shear
moduli~\cite{goodrich:14,goodrich:14b} in finite systems near jamming all scale the same way; the
$SO(3)$ rotational symmetry of isotropic jamming induces a symmetry in the 
various elastic moduli, all of which vanish linearly with $\Delta Z$ as $\Delta Z \to 0$ and $N \to \infty$.

A convenient way to see this symmetry is by viewing the elastic constant tensor  $C_\abcd$
as a linear transformation from the six-dimensional space of strain tensors
to the six-dimensional space of stress tensors (explicitly implemented,
for example, in Mandel notation). The resulting $6 \times 6$ Mandel stiffness matrix has 6 eigenvalues; 
at the isotropic jamming transition, one of these is nonzero while the remaining 5 vanish in the thermodynamic limit.  The eigenvector of the Mandel matrix corresponding to the nonzero eigenvalue corresponds to the bulk modulus, while the 5 eigenvectors corresponding to the vanishing eigenvalues are related to the shear moduli.  As the system is compressed above the isotropic jamming transition, the 5 eigenvectors remain the same, increasing linearly with $\Delta Z$.

\begin{figure}[!tb]
\centering
\includegraphics[width=0.8\columnwidth]{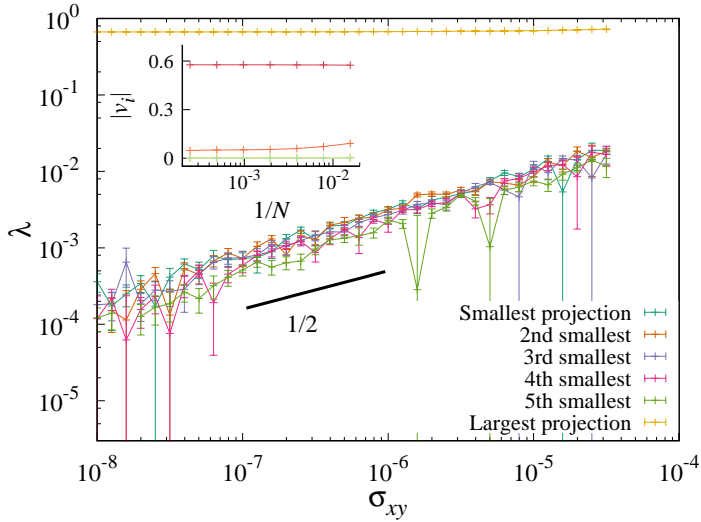}
\caption{Eigenvalues of the Mandel stiffness matrix $\Celas$, as a function of the stress along the shear, $\sigma_{xy}$, for $N=2048$. The six eigenvalues are
sorted according to their trace $\sigma_\ab$, reflecting the component of bulk
compression. Only one elastic component jumps at jamming, corresponding to
a combination of the bulk modulus and $C_{xyxy}$. The common quantitative scaling of the five
smaller eigenvalues suggests an emergent $SO(3)$ invariance, and the
exponent of 1/2 suggests that their scaling is the same as for isotropic jamming. 
In the inset we show the independent components of $X$, the eigenvector
associated to the largest eigenvalue $\lambda_X$, as a function of $1/N$.
From top to bottom, the data represent $\Tr(X)/3$,
$X_{xy}$, $X_{yy}-\Tr(X)/3$, $X_{zz}-\Tr(X)/3$, $X_{xz}$, $X_{yz}$.
Only the first two are distinguishable ($\Tr(X)/3$ and
$X_{xy}$); the remaining are all very close to zero.
}
\label{fig:cik}
\end{figure}

Figure~(\ref{fig:cik}) indicates that this same scenario appears 
to arise during shear jamming, despite the fact that shear breaks 
rotation invariance. Measuring the eigenvalues of the Mandel matrix 
transformation as in Ref.~\cite{roux:08b}, we find that
the constant values of both $B$ and 
$C_{xyxy}$ as the shear jamming transition is approached from above ($\sigma_{xy} \to 0^+$) both reflect the behavior of a {\em single} eigenvalue
of this transformation. Thus $C_\abcd = B' X_\ab X_\cd$ for some symmetric
tensor $X$ given by the corresponding eigenvector; the components of $X$ 
are described in the inset of Fig.~(\ref{fig:cik}). Furthermore, the
other five eigenvalues vanish at the shear-jamming transition linearly with $\Delta Z$, just as for
isotropic jamming. These results are consistent with earlier findings by Peyneau and Roux~\cite{roux:08b}, 
who found that one eigenvalue was much larger than the others in stressed systems close to the jamming transition.
In other words, both isotropic jamming and shear jamming select a distinguished direction $X$ in the six-dimensional
space of elastic moduli.  In the case of isotropic jamming, this direction projects entirely onto the bulk modulus, but for shear jamming it 
projects onto both $B$ and $C_{xyxy}$.  In both cases, we observe an emergent $SO(3)$ symmetry in the
remaining five dimensions.

Note from the inset to Figure~\ref{fig:cik} that by far the dominant contribution to $X$ comes from the bulk modulus, with only a small 
contribution from $C_{xyxy}$.  This behavior is reflected in the values of $B$ and $C_{xyxy}$ at the transition; while clearly nonzero, the magnitude 
of $C_{xyxy}$ is roughly an order of magnitude smaller than the value of $B$.  The small value of $C_{xyxy}$ and its small contribution to $X$ may be a 
result of the fact that the packing is frictionless.  Frictional packings may have much larger values of $C_{xyxy}$ relative to $B$ and much higher 
anisotropies in their structure.

Our results imply that the distinguishing hallmark of the jamming transition, whether isotropic or shear-induced, is to break symmetry 
in six-dimensional stress space below the jamming transition by picking out one distinguished direction at the jamming transition, 
leaving $SO(3)$ symmetry in the remaining five directions. 
{The scalings of the elastic constants are thus defined by the eigenvector $X$. Moduli that project onto $X$ exhibit a discontinuity at jamming, whereas
those in the subspace orthogonal to $X$ grow proportionally to $\Delta Z$.}
This suggests that the scaling
theory for shear jamming will be the same as for regular jamming~\cite{goodrich:16}, ``rotated"
in this six-dimensional space, and that the distinguished role of pressure
in scaling near isotropic jamming will be replaced in shear-jamming by that of the stress along the tensor
direction $X$.

\section{Discussion}
In summary, 
we find that shear-jammed states are different from isotropically-jammed states. This is due to the anisotropy induced by shear, even for frictionless sphere packings 
in the thermodynamic limit.  Shearing induces long-ranged bond correlations; as a result, the shear stress, $\sigma_{xy}$, is proportional to the pressure with a prefactor that does not depend on $N$ in the thermodynamic limit.  This leads to modified 
scaling exponents for $\sigma_{xy}$ and the elastic constant $C_{xyxy}$, so that $C_{xyxy}$, like the bulk modulus, jumps discontinuously at the onset of shear jamming. 

However, the differences between shear-jammed and isotropically-jammed states, although striking, are more superficial than fundamental. The discontinuous behavior of $C_{xyxy}$ 
in shear jamming is due to a rotation of the dominant eigenvector $X$ of the stiffness matrix. This rotation seems to preserve an emergent $SO(3)$ symmetry in the 5-dimensional 
subspace orthogonal to $X$. 
{Moduli exhibiting discontinuities at the jamming transition have nonzero projections onto the vector $X$, whereas elastic moduli in the orthogonal subspace scale linearly with $\Delta Z$.
The orientation of $X$ in elastic constant space depends on the loading history by which the system was jammed.}
An interesting question is whether this symmetry emerges for any combination of strains leading to jamming \cite{roux:08b} {, and what is the relation between $X$ and the applied strain}.

Note that there are other potential ways of creating shear-jammed states. One could start with an isotropically-jammed state, apply a shear stress and then decrease the pressure. This appears to lead to similar results~\cite{roux:08b}, although a more careful examination of behavior near the jamming transition is needed.
Alternatively, one could start with unjammed states at lower density and apply simultaneous shear and compression until the system jams.  With our protocol of starting with isotropically 
unjammed systems near the onset of jamming, and then straining the system until it shear jams, it becomes more difficult to create shear-jammed states with increasing system size because one 
must start with isotropically unjammed states that are closer to the onset of jamming. { The difficulty of preparing shear-jammed states is therefore an artifact of our protocol, and our results show that shear-jammed configurations in frictionless packings are well-defined in 
the thermodynamic limit.}

\appendix
\section{. Stress Correlation functions}
The scalings of $p$, $\sigma_{xy}$, $\sigma_{xz}$, $B$, $C_{xyxy}$ and $C_{xzxz}$ can be understood microscopically in terms 
of the behavior of spatial bond correlations.
We can express the average deviatoric squared stresses $\tilde{\sigma}^2$, $p^2$, $\sigma_{xy}^2$ and $\sigma_{xz}^2$ in 
terms of associated stress correlation functions $\C,\Cp,\Cxy$ and $\Cxz$.

In order to do so, we extend the correlation function $\C$ defined in reference \cite{goodrich:16} to take into account the anisotropy
caused by the shear.
The stress tensor is defined through
\begin{equation}
 \sigma_{\alpha\beta} = - \frac{1}{V} \sum_k^{\Nb} \bk \rka \rkb\,,
\end{equation}
where the index $k$ indicates a contact (bond) between two particles, $\Nb$ is the number of bonds, $\vec{r}^{(k)}=\frac{\rk}{|\vec{r}^{(k)}|}$ is the separation 
between the two touching particles, and $\bk=f^{(k)} |\vec{r}^{\,(k)}|$,  where $f^{(k)}$ is the force of bond $k$. 
The product between generic components of the stress tensor is
\begin{align}\label{eq:sabscd}
 \sigma_\ab\sigma_\cd &= \frac{1}{V^2}\sum_{k,k'} \bk\bkp\rka\rkb\rkp_\gamma\rkp_\delta =  \\\nonumber
                      &= \frac{1}{V^2}\sum_{k} {\bk}^2\rka\rkb\rk_\gamma\rk_\delta + \\\nonumber
                      &+\frac{1}{V^2}\sum_{k\neq k'}\bk\bkp\rka\rkb\rkp_\gamma\rkp_\delta =\\\nonumber
\end{align}
Taking out a factor $N_b$, both terms can be seen as an average over the bonds,
\begin{align}
 \sigma_\ab\sigma_\cd &= \frac{\Nb}{V^2}\left\langle b^2\,\rka\rkb\rk_\gamma\rk_\delta\right\rangle+\\\nonumber
                      &+ \frac{\Nb}{V^2}\left\langle \sum_{k\neq0} \bz\bk\,\rza\rzb\rk_\gamma\rk_\delta\right\rangle\,.
\end{align}
From the second term in the right hand side (r.h.s.) we can define a correlation function 
\begin{equation}
  \Cabcd  = \left\langle \sum_{k\neq0} \bz\bk \left[\rza \rzb \rkc \rkd \right] \delta\left(\left[\xz-\xk\right]-\vec x\right)\right\rangle\,,
\end{equation}
where $\xk$ is the position of bond $k$.

The correlation function $\Cabcd$ is related to the stress through
\begin{equation}\label{eq:sabcdC}
  \sigma_\ab\sigma_\cd = \frac{\Nb}{V^2}\left\langle b^2\,\rka\rkb\rk_\gamma\rk_\delta\right\rangle + \frac{\Nb}{V^2}\int d^dx \,\Cabcd\,.
\end{equation}
 
With an analogous procedure it is possible to define a wide variety of correlation functions, 
\footnote{For example one could be interested in the correlators obtained by expanding the traceless stress tensor
or the pressure.}
each with a relation that connects it to the stress.

In this instance we are interested in the correlation functions
\begin{align}
\label{eq:C}
 \C  &= \left\langle \sum_{k\neq0} \bz\bk \left[(\rz\cdot \rk)^2 -1/d\right] \delta\left(\left[\xz-\xk\right]-\vec x\right)\right\rangle\,,\\
\label{eq:Cp}
 \Cp  &= \left\langle \sum_{k\neq0} \bz\bk \left[\frac{1}{d^2}\right]\delta\left(\left[\xz-\xk\right]-\vec x\right)\right\rangle\,,\\
\label{eq:Cxy}
 \Cxy &= \left\langle \sum_{k\neq0} \bz\bk \left[\rzx \rzy \rkx \rky  \right] \delta\left(\left[\xz-\xk\right]-\vec x\right)\right\rangle\,,\\
\label{eq:Cxz}
 \Cxz &= \left\langle \sum_{k\neq0} \bz\bk \left[\rzx \rzz \rkx \rkz  \right] \delta\left(\left[\xz-\xk\right]-\vec x\right)\right\rangle\,,
 \end{align}
that are related to the stress through
\begin{align}
\label{eq:stildeC}
\tilde{\sigma}^2	&= \frac{\Nb}{V^2}\left\langle b^2\right\rangle\frac{d-1}{d} + \frac{\Nb}{V^2}\int d^dx \,\C\,,\\[1ex]
\label{eq:pC}
p^2         		&= \frac{\Nb}{V^2d^2}\left\langle b^2\right\rangle           + \frac{\Nb}{V ^2}\int d^dx \,\Cp\,,\\[1ex]
\label{eq:sxyC}
\tilde\sigma_{xy}^2 	&= \frac{\Nb}{V^2}\left\langle b^2\,\hat{r}_x^2\,\hat{r}_y^2 \right\rangle + \frac{\Nb}{V ^2}\int d^dx \,\Cxy\,,\\[1ex]
\label{eq:sxzC}
\tilde\sigma_{xz}^2 	&= \frac{\Nb}{V^2}\left\langle b^2\,\hat{r}_x^2\,\hat{r}_z^2 \right\rangle + \frac{\Nb}{V ^2}\int d^dx \,\Cxz\,,
\end{align}

The first term in the r.h.s. of Eqs. \eqref{eq:sabcdC}, \eqref{eq:stildeC}, \eqref{eq:pC}, \eqref{eq:sxyC}, \eqref{eq:sxzC} is of order $\left<b^2\right>/N$, while the second 
term depends on the integral of the correlation function.  Let $\mathcal{O}^2$ represent the l.h.s of any of these equations. 
If {the corresponding correlation function is short-ranged then the integral is proportional to $\left<b^2\right> N^0$, and $\mathcal{O}^2\sim \left<b^2\right>/N \sim p^2/N$.
However, if the correlation function is long-ranged, then the integral is proportional to $\left<b^2\right> N$, and 
\begin{equation}\label{eq:opn}
\mathcal{O}^2 \sim \left<b^2\right> N^0 \sim p^2 N^0
\end{equation}
}

\begin{figure}[!tbh]
\centering
\includegraphics[width=0.8\columnwidth]{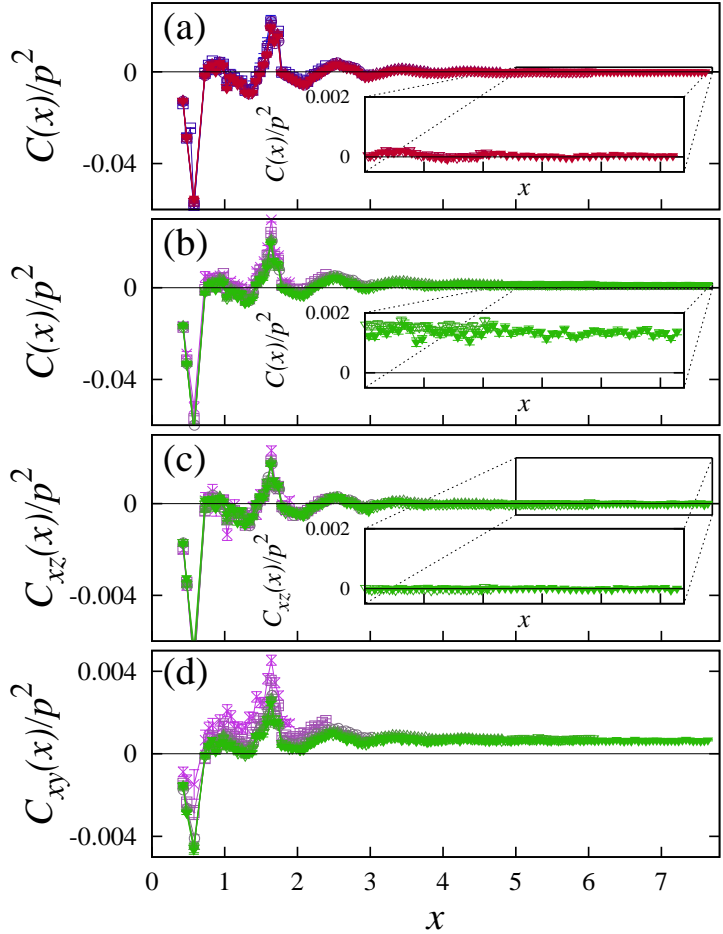}
\caption{Bond correlations $\C,\Cxy$ and $\Cxz$, normalized with the square pressure $p^2$, as a function of the distance $x$ between bonds, for all system sizes at $\sigma_{xy}=10^{-8}$.
The integrals of the correlation functions are related with the square stresses through Eqs.~\eqref{eq:stildeC}, \eqref{eq:pC}, 
\eqref{eq:sxyC}, \eqref{eq:sxzC}: if the integral is extensive (i.e. the correlation in long-ranged)
then the related squared stress is of order ${p^2}$. If it is intensive (short-range correlations), the {squared} stress is of order $p^2/N$.
(a) Correlation $\C$ in isotropically-jammed packings. It is short-ranged {(inset)}, so $\tilde{\sigma}^2\sim {p^2}/N$ in such systems.
(b) Correlation $\C$ in shear-jammed packings. It does not decay to zero {(inset)}, so $\tilde{\sigma}^2\sim p^2$ in such systems.
(c) Correlation $\Cxz$ in shear-jammed packings. There are no long-ranged correlations orthogonal to the direction of shear jamming {(inset)}, so $\sigma_{xz}\sim {p^2}/N$.
(d) Correlation $\Cxy$ in shear-jammed packings. The long-distance correlations decay to a positive constant, so $\sigma_{xy}\sim p^2$.
The insets depict a zoom of the same data of the larger plot, so that the differences between long-ranged correlation functions and short-ranged ones are more visible.
All the insets have the same range.
}
\label{fig:corr}
\end{figure}

In figure \ref{fig:corr}{(a) and (b)} we show that the correlation function $\C$ is short-ranged for isotropically-jammed states, and long-ranged for shear-jammed 
states: shear-jamming induces long-range correlations in the system, which lead to a non-zero deviatoric stress.
The two bottom plots of Fig.~\ref{fig:corr} show that the correlation $\Cxy$ along the shear is 
long-ranged, whereas in the orthogonal direction $\Cxz$ is 
short-ranged, explaining why $\sigma_{xy}$ and $\sigma_{xz}$ scale differently.

The long-ranged nature of $\Cxy$ leads to the scaling $\sigma_{xy}^2 \sim p^2$, consistent with Figure~\ref{fig:dZ-p}. 
This in turn leads to the prediction that the scaling exponents for $C_{xyxy}$ and $B$ are the same, consistent with Figure~\ref{fig:G}.

%%%%%%%%%%%%%%%%%%
%ACKNOWLEDGEMENTS%
%%%%%%%%%%%%%%%%%%
\begin{acknowledgements}
The authors owe a great intellectual debt to Leo Kadanoff.  
His demonstration of how scaling arguments can be used to understand and categorize the universality of physical phenomena was an inspiration for many of the ideas in this paper.  
His catholic taste in choosing problems, his success in developing simple models to understand complex phenomena, 
and his complete lack of snobbishness in deciding what problems were important allowed theory and experiment to work together to make progress throughout many areas of science.  
We dedicate this paper in his memory.

We thank  Valerio Astuti, Eric DeGiuli, Edan Lerner and Pierfrancesco Urbani for interesting discussions.
This work was funded by the Simons Foundation for the collaboration ``Cracking the Glass Problem" (454945 to A.J.L. for A.J.L. and J.P.S.,
348126 to S.R.N. for S.R.N., and 454935 to G. Biroli for M.B.-J.), the National Science Foundation (DMR-1312160 for J.P.S.), 
the ERC grant NPRGGLASS (279950 for M.B.-J.), and the US Department of Energy, Office of Basic Energy Sciences, 
Division of Materials Sciences and Engineering under Award DE-FG02-05ER46199 (C.P.G.).
M.B.-J. was also supported by MINECO, Spain, through  the  research  contract N$^o$. FIS2012-35719-C02,
and by the  FPU  program  (Ministerio  de  Educaci\'on,  Spain).
\end{acknowledgements}


\begin{thebibliography}{}
\bibitem{liu:10} A. J. Liu and S. R. Nagel, Annu. Rev. Condens. Matter Phys. 2010. 1:347–69.
\bibitem{xu:07} Xu, N., Wyart, M., Liu, A. J. and Nagel, S. R. Excess vibrational modes and the boson peak in model glasses. Phys. Rev. Lett. \textbf{98}, 175502 (2007).
\bibitem{wyart:08} Wyart, M., Liang, H., Kabla, A. and Mahadevan, L. Elasticity of floppy and stiff random networks. Phys. Rev. Lett. \textbf{101}, 215501 (2008).
\bibitem{phillips79} Phillips, J. C. Topology of covalent non-crystalline solids I: Short-range order in chalcogenide alloys. J. Non-Cryst. Solids \textbf{34}, 153-181 (1979).
\bibitem{phillips:81} Phillips, J. C. Topology of covalent non-crystalline solids II: Medium-range order in chalcogenide alloys and A Si (Ge). J. Non-Cryst. Solids \textbf{43}, 37-77 (1981).
\bibitem{boolchand:05} Boolchand, P., Lucovsky, G., Phillips, J. C. and Thorpe, M. F. Self-organization and the physics of glassy networks. Phil. Mag. \textbf{85}, 3823-3838 (2005).
\bibitem{song:08} Song, C., Wang, P. and Makse, H. A. A phase diagram for jammed matter. Nature \textbf{453}, 629-632 (2008).
\bibitem{henkes:10} Henkes, S., van Hecke, M. and van Saarloos, W. Critical jamming of frictional grains in the generalized isostaticity picture. Europhys. Lett. \textbf{90}, 14003 (2010).
\bibitem{papanikolaou:13} Papanikolaou, S., O'Hern, C. S. and Shattuck, M. D. Isostaticity at frictional jamming. Phys. Rev. Lett. 110, 198002 (2013).
\bibitem{zhang:09} Zhang, Z. et al. Thermal vestige of the zero-temperature jamming transition. Nature \textbf{459}, 230-233 (2009).
\bibitem{donev:04} Donev, A. et al. Improving the density of jammed disordered packings using ellipsoids. Science \textbf{303}, 990-993 (2004).
\bibitem{zeravcic:09} Zeravcic, Z., Xu, N., Liu, A. J., Nagel, S. R. and van Saarloos, W. Excitations of ellipsoid packings near jamming. Europhys. Lett. \textbf{87}, 26001 (2009).
\bibitem{mailman:09} Mailman, M., Schreck, C. F., O'Hern, C. S. and Chakraborty, B. Jamming in systems composed of frictionless ellipse-shaped particles. Phys. Rev. Lett. \textbf{102}, 255501 (2009).
\bibitem{goodrich:14c} C.Goodrich, A.Liu and S.Nagel, Nature Physics \textbf{10}, 578-581 (2014).
\bibitem{somfai:07} E. Somfai, M. van Hecke, W. G. Ellenbroek, K. Shundyak, and W. van Saarloos, Phys. Rev. E {\bf 75}, 020301(R) (2007).
\bibitem{shundyak:07} K. Shundyak, M. van Hecke and W. van Saarloos, Phys. Rev. E {\bf 75}, 010301(R) (2007).
 \bibitem{bi:11}  D. Bi, J. Zhang, B. Chakraborty and R. P. Behringer, Nature \textbf{480}, 355–358 (2011).
 \bibitem{bertrand:16} T. Bertrand, R. P. Behringer, B. Chakraborty, C. S. O'Hern, M. D. Shattuck, Phys. Rev. E \textbf{93}, 012901 (2016).
 \bibitem{imole:13} O.I. Imole, N. Kumar, V. Magnanimo and S. Luding, Kona Powder and Particle Journal \textbf{30}, 84-108 (2013).
 \bibitem{vinutha:16} H.A. Vinutha and S. Sastry, Nat. Phys. (2016), arXiv:1510.00962, DOI:10.1038/nphys3658. 
 \bibitem{vinutha:16b} H.A. Vinutha and S. Sastry, in preparation. 
 \bibitem{roux:08a} P.-E. Peyneau, J.-N. Roux, Phys. Rev. E \textbf{78}, 011307 (2008).
 \bibitem{roux:08b} P.-E. Peyneau, J.-N. Roux, Phys. Rev. E \textbf{78}, 041307 (2008). 
 \bibitem{goodrich:16} C. P. Goodrich, A. J. Liu and J. P. Sethna, Proc. Nat. Ac. Sci. \textbf{113}, 35, 9745–9750 (2016).
 \bibitem{wyart:05} M. Wyart, Annales de Physique \textbf{30}, 3 (2005), DOI:10.1051/anphys:2006003.
 \bibitem{wyart:11} M. Wyart, Microgels: Synthesis, Properties and Applications, edited by A. Fernandez, J. Mattsson, H.M. Wyss, D.A. Weitz, Wiley \& Sons, Weinheim (2011) p. 195-206, arXiv:0806.4653.
 \bibitem{zaccone:14} A. Zaccone and E.M. Terentjev,  J. Appl. Phys. \textbf{115}, 033510 (2014); http://dx.doi.org/10.1063/1.4862403.
 \bibitem{vitelli:10} V. Vitelli, N. Xu, M. Wyart, A.J. Liu and S.R. Nagel, Phys. Rev. E \textbf{81}, 021301 (2010).
 \bibitem{bitzek:06} E.  Bitzek,  P.  Koskinen,  F.  G\"ahler,  M.  Moseler,  and P. Gumbsch, Phys. Rev. Lett. \textbf{97}, 170201 (2006).
 \bibitem{ohern:03} C.S. O'Hern, L.E. Silbert, A.J. Liu and S.R. Nagel, Phys. Rev. E \textbf{68}, 011306(2003).
 \bibitem{madadi:04} M. Madadi, O. Tsoungui, M. L\"atzel and S. Luding, Int. J. Sol. Str. \textbf{41}, 9, 2563, (2004).
 \bibitem{goodrich:12} C. P. Goodrich, A. J. Liu and S. R. Nagel, Phys. Rev. Lett. \textbf{109}, 9, 095704 (2012).
 \bibitem{goodrich:14} C.P. Goodrich, S. Dagois-Bohy, B. P. Tighe, M. van Hecke, A.J. Liu, S.R. Nagel, Phys. Rev. E \textbf{90}, 022138 (2014).
 \bibitem{goodrich:14b} C.P. Goodrich, A.J. Liu, S.R. Nagel, Phys. Rev. E \textbf{90}, 022201 (2014).
\end{thebibliography}
\end{document}